\documentclass{article}
\usepackage{waspaa19,amsmath,graphicx}
\usepackage{hyperref, url}
\usepackage{balance,cite}

\newcommand{\defeq}{\stackrel{\tiny{\mathrm{def}}}{=}}

\DeclareMathOperator*{\argmin}{argmin}

\makeatletter
\def\hlinewd#1{%
  \noalign{\ifnum0=`}\fi\hrule \@height #1 \futurelet
   \reserved@a\@xhline}
\makeatother

\title{Universal Sound Separation}
\name{
\parbox{\linewidth}{
\centering
Ilya Kavalerov$^{1,2}$\sthanks{Work done during an internship at Google.},
Scott Wisdom$^{1}$,
Hakan Erdogan$^{1}$,
Brian Patton$^{1}$,
\\
Kevin Wilson$^{1}$,
Jonathan Le Roux$^{3}$,
John R. Hershey$^{1}$
}}
\address{$^{1}$Google Research, Cambridge MA \\ $^2$Department of Electrical and Computer Engineering, UMD \\  $^{3}$Mitsubishi Electric Research Laboratories (MERL), Cambridge, MA}

\begin{document}
\ninept
\maketitle
\begin{abstract}
Recent deep learning approaches have achieved impressive performance on speech enhancement and separation tasks. However, these approaches have not been investigated for separating mixtures of arbitrary sounds of different types, a task we refer to as {\it universal sound separation}, and it is unknown how performance on speech tasks carries over to non-speech tasks. To study this question, we develop a dataset of mixtures containing arbitrary sounds, and use it to investigate the space of mask-based separation architectures, varying both the overall network architecture and the framewise analysis-synthesis basis for signal transformations. These network architectures include convolutional long short-term memory networks and time-dilated convolution stacks inspired by the recent success of time-domain enhancement networks like ConvTasNet. For the latter architecture, we also propose novel modifications that further improve separation performance. In terms of the framewise analysis-synthesis basis, we explore both a short-time Fourier transform (STFT) and a learnable basis, as used in ConvTasNet.  For both of these bases, we also examine the effect of window size.  In particular, for STFTs, we find that longer windows (25-50 ms) work best for speech/non-speech separation, while shorter windows (2.5 ms) work best for arbitrary sounds. For learnable bases, shorter windows (2.5 ms) work best on all tasks. Surprisingly, for universal sound separation, STFTs outperform learnable bases.  Our best methods produce an improvement in scale-invariant signal-to-distortion ratio of over 13 dB for speech/non-speech separation and close to 10 dB for universal sound separation.
\end{abstract}
\vspace{-5pt}
\section{Introduction}
\label{sec:intro}
A fundamental challenge in machine hearing is that of selectively listening to different sounds in an acoustic mixture.  Extracting estimates of each source is especially difficult in monaural recordings where there are no directional cues. 
Recent advances have been made in solving monaural speech enhancement and speech separation problems in increasingly difficult scenarios, thanks to deep learning methods \cite{lu2013speech,Weninger2014RNN,xu2014experimental,erdogan2015phase,weninger2015speech,Hershey2016ICASSP03,Isik2016Interspeech09,Yu2017PIT,kolbaek2017uPIT,Wang2017Overview,Wang2018ICASSP04Alternative}.  However, separation of arbitrary sounds from each other may still be considered a ``holy grail'' of the field.  

In particular, it is an open question whether current methods are best suited to learning the specifics of a single class of sounds, such as speech, or can learn more general cues for separation that can apply to mixtures of arbitrary sounds.  In this paper, we propose a new universal sound separation task, consisting of mixtures of hundreds of types of sound.  We show that the best methods are surprisingly successful, producing an average improvement of almost 10 dB in scale-invariant signal-to-distortion ratio (SI-SDR) \cite{leroux2018sdr}. 

Previous experiments have focused mainly on scenarios where at least one of the target signals to be separated is speech. In speech enhancement, the task is to separate the relatively structured sound of a single speaker from a much less constrained set of non-speech sounds. For separation of multiple speakers, the state of the art has progressed from speaker-dependent separation \cite{huang2014deep}, where models are trained on individual speakers or speaker combinations, to speaker-independent speech separation \cite{Hershey2016ICASSP03,Isik2016Interspeech09,Yu2017PIT}, where the system has to be flexible enough to separate unknown speakers.  In particular, ConvTasNet is a recently proposed model \cite{luo2018tasnet} that uses a combination of learned time-domain analysis and synthesis transforms with a time-dilated convolutional network (TDCN), showing significant improvements on the task of speech separation relative to previously state-of-the-art models based on short-time Fourier transform (STFT) analysis/synthesis transforms and long short-term memory (LSTM) recurrent networks.  Despite this progress, it is still unknown how current methods perform on separation of arbitrary types of sounds.  The fact that human hearing is so adept at selective listening suggests that more general principles of separation exist and can be learned from large databases of arbitrary sounds.  

\begin{figure*}[tbh]
\centering
\includegraphics[width=.95\linewidth]{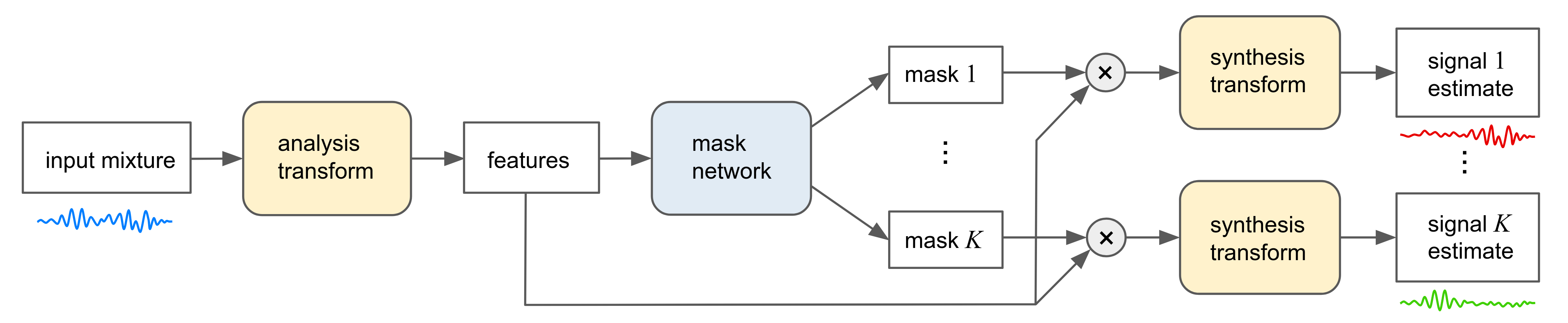}
\vspace{-2mm}
\caption{Architecture for mask-based separation experiments. We vary the mask network and analysis/synthesis transforms. }
\label{fig:network}
\vspace{-10pt}
\end{figure*}

This paper provides four contributions.
    First, we investigate the universal sound separation problem in depth for the first time, by constructing a dataset of mixtures containing a wide variety of different sounds. 
    Second, we evaluate ConvTasNet on both speech/non-speech separation and universal sound separation tasks for the first time.
    Third, we provide a systematic comparison of different combinations of masking network architectures and analysis-synthesis transforms, optimizing each over the effect of window size.  %
    Finally, we propose novel variations in architecture, including alternative feature normalization, improved initialization, longer range skip-residual connections, and iterative processing that further improve separation performance on all tasks.

\vspace{-5pt}
\section{Prior Work}
\vspace{-2.5pt}
\label{sec:prior}

A variety of networks have been successfully applied to two-source separation problems, including LSTMs and bidirectional LSTMs (BLSTMs) \cite{Weninger2014RNN,erdogan2015phase}, U-Nets \cite{unet_spot}, Wasserstein GANs \cite{gan}, and fully convolutional network (FCN) encoder-decoders followed by a BLSTM \cite{plumbley_latest}.
For multi-source separation, a variety of architectures have been used that directly generate a mask for each source, including BLSTMs \cite{Hershey2016ICASSP03,kolbaek2017uPIT}, CNNs \cite{plumbley}, DenseNets
followed by an LSTM \cite{mmdenselstm}, 
separate encoder-decoder networks for each source \cite{cdae},
joint one-to-many encoder-decoder networks with o decoder per source \cite{onetomany}, and TDCNs with learnable analysis-synthesis basis \cite{luo2018tasnet}.
Our models are most similar to \cite{kolbaek2017uPIT} and \cite{luo2018tasnet}.  Networks that perform source separation in an embedding space instead of in the time-frequency domain, such as deep clustering \cite{Hershey2016ICASSP03,Wang2018ICASSP04Alternative}, have also been effective at separation tasks, but we leave exploration of those methods for future work.

Previous source separation work has focused on speech enhancement and speech separation
\cite{gan, chime2,Hershey2016ICASSP03,attractor}.
Small datasets used for the non-speech multi-source separation setting have included distress sounds from DCASE 2017 \cite{plumbley}, and speech and music in SiSEC-2015 \cite{cdae,plumbley_latest}.  Singing voice separation has focused on vocal and music instrument tracks 
\cite{dpcl_music, unet_spot}.

To our knowledge, the work introduced here is the first to investigate separation of arbitrary real-world sounds sourced from a large number of sound classes.

\vspace{-5pt}
\section{Models}
\label{sec:models}
\vspace{-2.5pt}

We use mask-based separation systems driven by deep neural networks, and we experiment with combinations of two different network architectures and two different analysis-synthesis bases. All masking networks use a sigmoid activation to predict a real number in $[0, 1]$ to modulate each basis coefficient.

\vspace{-7.5pt}
\subsection{Masking network architectures}
The first masking network we use consists of 14 dilated 2D convolutional layers, a bidirectional LSTM, and two dense layers, which we will refer to as a convolutional-LSTM-dense neural network (CLDNN). The CLDNN is based on a network which achieves state-of-the-art performance on CHiME2 WSJ0 speech enhancement \cite{wilson2018exploring} and strong performance on a large internal dataset \cite{wisdom2018differentiable}.

Our second masking network is a TDCN inspired by ConvTasNet \cite{luo2018tasnet}. We employ the same parameters as the best noncausal model reported by %
\cite{luo2018tasnet}.
We also consider an improved version of ConvTasNet's TDCN masking network, which we refer to as ``improved TDCN'' (TDCN++). This new architecture includes three improvements to the original ConvTasNet network. First, global layer normalization within the TDCN,
which normalizes over all features and frames, is replaced with a feature-wise layer normalization over frames. This is inspired by cepstral mean and variance normalization used in automatic speech recognition systems. Second, we add longer-range skip-residual connections from earlier repeat inputs to later repeat inputs after passing them through dense layers. This presumably helps with gradient flow from layer to layer during training. Third, we add a learnable scaling parameter after each dense layer.
The scaling parameter for the second dense layer in each convolutional block -- which is applied right before the residual connection -- is initialized to an exponentially decaying scalar equal to $0.9^L$, where $L$ is the layer or block index. This initial scaling contributes to better training convergence by first learning the contributions of the bottom layers, similar to layer-wise training, and then easily adjusting the scale of each block's contribution through the learnable scaling parameter. This initialization is partly inspired by ``Fixup'' initialization in residual networks \cite{zhang2019fixup}.

A third network variant we consider is an iterative improved TDCN network (iTDCN++), in which the signal estimates from an initial mask-based separation network serve as input, along with the original mixture, to a second separation network.   This architecture is inspired by \cite{Isik2016Interspeech09}, in which a similar iterative scheme with LSTM-based networks led to significant performance improvements.   In our version, both the first and second stage networks are identical copies of the TDCN++ network architecture, except for the inputs and parameters.  In the second stage, the noisy mixture and initial signal estimates are transformed by the same basis (STFT or learned) prior to concatenation of their coefficients.     Because, with two iterations, the network is twice as deep as a single-stage TDCN++, we also include a twice deeper TDCN++ model (2xTDCN++) for comparison.  

\vspace{-7.5pt}
\subsection{Analysis-synthesis bases}
Whereas earlier mask-based separation work had used STFTs as the analysis-synthesis basis due to the sparsity of many signals in this domain,  ConvTasNet \cite{luo2018tasnet} uses a learnable analysis-synthesis basis. The analysis transform is a framewise basis analogous to the STFT, and can also be described as a 1D convolution layer where the kernel size is the window size, the stride is the hop length, and the number of filters is the number of basis vectors.    A ReLU activation is applied to the analysis coefficients before processing by the mask network. The learnable synthesis transform can be expressed as a transposed 1D convolution and operates similarly to an inverse STFT, where a linear synthesis basis operates on coefficients to produce frames which are overlap-added to form a time-domain signal. Unlike an STFT, this learnable basis and its resulting coefficients are real-valued.   

The original work \cite{luo2018tasnet} found that ConvTasNet performed best with very short (2.5 ms) learnable basis functions.  However, this window size is an important parameter that needs to be optimized for each architecture, input transform, and data type.  We therefore compare a learnable basis with STFT as a function of window size, in combination with CLDNN and TDCN masking networks. All models apply mixture consistency projections to their outputs \cite{wisdom2018differentiable}, which ensure the estimated sources add up to the input mixture. Note that the TDCN with STFT basis is a novel combination that, as we show below, performs best on the universal separation task.

\vspace{-5pt}
\section{Experiments}
\label{sec:exp}
\vspace{-2.5pt}

\begin{figure*}[tp]
\centering

\begin{minipage}[b]{1.0\linewidth}
\centerline{\includegraphics[width=1.0\linewidth]{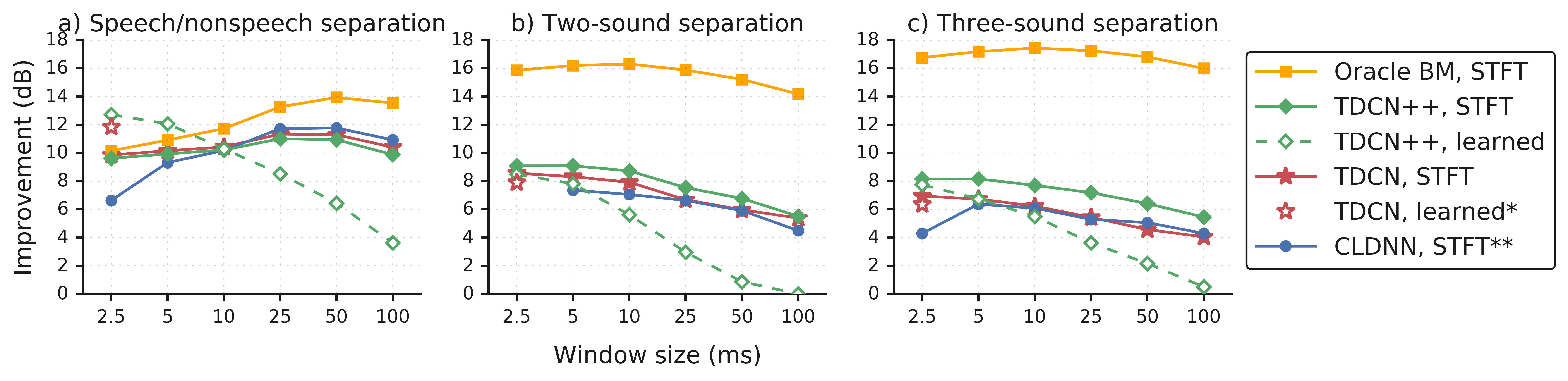}}
\vspace{-10pt}
 \caption{Mean SI-SDR improvement in dB on the test set as a function of basis window size in ms, using different combinations of network architectures and bases, on a) speech/non-speech separation, b) two-sound universal separation, and c) three-sound universal separation. Systems * and ** come from \cite{luo2018tasnet} and \cite{wilson2018exploring,wisdom2018differentiable}, respectively. ``Oracle BM'' corresponds to an oracle binary STFT mask, a theoretical upper bound on our systems' performance.  Note the CLDNN STFT failed to converge for 2.5 ms windows on two-sound separation and is omitted.
 }
\label{fig:sisnr_vs_stftwin}
\vspace{-12.5pt}
\end{minipage}
\end{figure*}

In this section, we describe the construction of a dataset for universal sound separation and apply the described combinations of masking networks and analysis-synthesis bases to this task.

\subsection{Dataset construction}

We define a \emph{universal sound separation} task designed to have tremendous variability.  To build a dataset for this task, we used the Pro Sound Effects Library database \cite{prosound}, which contains an encyclopedic sampling of movie production recordings, including crawling insects, animal calls, creaking doors, construction noises, musical instruments, speech, composed music, and artificial sounds (e.g., arcade game sounds). Ambience/environment tracks are excluded since they tend to include multiple overlapping sounds.

Three-second clips were extracted from the Pro Sound database and used to create single-channel mixtures. 
Each sound file was analyzed to identify the start of individual sound events, by detecting when the local
root-mean-squared power changed from below average to above average.
For each of these detected event times within a file, a three-second segment was extracted, where the center of the segment is equal to the detected event time plus a random uniform offset of up to half a second. Files that were shorter than 3 seconds were looped with a random delay of up to a second to create a three-second segment.

To create each three-second mixture clip, $K$ source clips were chosen from different sound files randomly and added together.
The data were partitioned by source file, with 70\% of the files used in the training set, %
 20\% in the validation set, and 10\% in the test set.  
 Overall, the source material for the training set consists of 11,797 audio files, along with 3,370 for the validation set, and 1,686 for the test set. 
In total, the two-source and three-source datasets each contain 14,049 training mixtures (11.7 hours), 3898 validation mixtures (3.2 hours), and 2074 test mixtures (1.7 hours). A recipe to recreate the dataset is publicly available \cite{supplemental_webpage}.

\begin{table*}[t]
\caption{Mean scale-invariant SDR improvement (dB) for speech/non-speech separation and two-source or three-source sound separation. Note that the bottom four TDCN networks below the thick line are twice as deep as the top four TDCN networks above the thick line.
}
\label{tab:divsource}
\centering
{\footnotesize
\begin{tabular}{|l|ccc|ccc|ccc|}
\hline
& \multicolumn{3}{c|}{Speech/non-speech separation}
& \multicolumn{3}{c|}{Two-source separation}
& \multicolumn{3}{c|}{Three-source separation}
\\
{\bf Masking network, basis}
& \begin{tabular}{@{}c@{}}{\bf Best} \\ {\bf win.\ size}\end{tabular}
& \begin{tabular}{@{}c@{}}{\bf Val.} \\ {\bf SI-SDRi}\end{tabular}
& \begin{tabular}{@{}c@{}}{\bf Test} \\ {\bf SI-SDRi}\end{tabular}
& \begin{tabular}{@{}c@{}}{\bf Best} \\ {\bf win.\ size}\end{tabular}
& \begin{tabular}{@{}c@{}}{\bf Val.} \\ {\bf SI-SDRi}\end{tabular}
& \begin{tabular}{@{}c@{}}{\bf Test} \\ {\bf SI-SDRi}\end{tabular}
& \begin{tabular}{@{}c@{}}{\bf Best} \\ {\bf win.\ size}\end{tabular}
& \begin{tabular}{@{}c@{}}{\bf Val.} \\ {\bf SI-SDRi}\end{tabular}
& \begin{tabular}{@{}c@{}}{\bf Test} \\ {\bf SI-SDRi}\end{tabular}
\\  \hline
CLDNN, STFT \cite{wilson2018exploring,wisdom2018differentiable}
& 50 ms & 11.9 & 11.8
& 5.0 ms & \phantom{1}7.8 & 7.4
& 5.0 ms & 6.7 & 6.4
\\ 
\hline
TDCN, learned \cite{luo2018tasnet}
& 2.5 ms & 12.6 & 12.5
& 2.5 ms & \phantom{1}8.5 & 7.9
& 2.5 ms & 6.8 & 6.4
\\
TDCN, STFT
& 25 ms & 11.5 & 11.3
& 2.5 ms & \phantom{1}9.4 & 8.6
& 2.5 ms & 7.6 & 7.0
\\  \hline
TDCN++, learned
& 2.5 ms & {\bf 12.7} & {\bf 12.7}
& 2.5 ms & \phantom{1}9.1 & 8.5
& 2.5 ms & 8.4 & 7.7
\\ 
TDCN++, STFT
& 25 ms & 11.1 & 11.0
& 2.5 ms & {\bf \phantom{1}9.9} & {\bf 9.1} 
& 5.0 ms & {\bf 8.8} & {\bf 8.2}
\\ \hlinewd{1pt}
2xTDCN++, learned
& 2.5 ms & 13.3 & 13.2
& 2.5 ms & \phantom{1}8.1 & 7.6
& 2.5 ms & 8.0 & 7.3
\\
2xTDCN++, STFT
& 25 ms & 11.2 & 11.1
& 5.0 ms & \phantom{1}9.3 & 8.3
& 5.0 ms & 9.0 & 8.0
\\ \hline
iTDCN++, learned
& 2.5 ms & {\bf 13.5} & {\bf 13.4}
& 2.5 ms & \phantom{1}9.3 & 8.7
& 2.5 ms & 8.1 & 7.4
\\ 
iTDCN++, STFT
& 25 ms & 11.6 & 11.5
& 2.5 ms & {\bf 10.6} & {\bf 9.8}
& 2.5 ms & {\bf 9.6} & {\bf 8.7}
\\
\hline
\end{tabular}
}
\vspace{-5pt}
\end{table*}

\subsection{Training and evaluation setup}
 All experiments are performed using TensorFlow \cite{tf_short}, trained with the Adam \cite{adam} optimizer 
with batch size 2 on a single NVIDIA Tesla V100 GPU.
Separation performance is measured using scale-invariant signal-to-distortion ratio improvement (SI-SDRi) \cite{Isik2016Interspeech09, leroux2018sdr}, which evaluates the fidelity of a signal estimate $\hat{s}$, represented as a vector, relative to the ground truth signal $s$ while accommodating a possible scale mismatch. SI-SDR is computed as
\begin{align}
    \text{SI-SDR}(s, \hat{s}) &\defeq  10 \log_{10} \frac{\| \alpha s\|^2}{\| \alpha s - \hat{s}\|^2},
\end{align}
where $ \alpha = \argmin_{a} \| a s - \hat{s}\|^2 = \langle s, \hat{s}\rangle /\|s\|^2$, and $\langle a,b \rangle$ denotes the inner product.  SI-SDRi is the difference between the SI-SDR of the estimated signal and that of the input mixture signal.
The sample rate for the mixtures was 16 kHz, and all STFTs use a square-root Hann window, where windowed frames are zero-padded to the next power of 2 above the window size.

We use a permutation-invariant loss to align network outputs with the reference sources during training, where the loss used for a gradient step on a batch is the minimum error across the set $S_K$ of all permutations of the $K$ estimated sources, compared to the fixed $K$ reference sources \cite{Hershey2016ICASSP03,Isik2016Interspeech09,Yu2017PIT}.
Although the cardinality of $S_K$ is $K!$, in our experiments $K\leq 3$ and this minimization did not lengthen training time significantly. Even for larger $K$, 
the time-consuming loss function computation can be first done in parallel for all pairs $(i,j), 1\leq i, j\leq K$, 
and the exhaustive search over permutations for the best combination is performed on the scores.

All networks use negative signal-to-noise ratio (SNR) as their training loss $f$ between time-domain reference source $y$ and separated source $\hat{y}$, defined as
\begin{equation}
    f(y, \hat{y})
    =
    -10
    \log_{10}
    \left(
        \frac
        {\sum_t y_t^2}
        {\sum_t (y_t - \hat{y}_t)^2}
    \right).
\end{equation}
Compared to negative SI-SDR used to train ConvTasNet \cite{luo2018tasnet}, this negative SNR objective has the advantage that the scale of separated sources is preserved and consistent with the mixture, which is further enforced by our use of mixture consistency layers \cite{wisdom2018differentiable}. Since we measure loss in the time domain, gradients are backpropagated through the synthesis transform and its overlap-add layer, so STFT consistency \cite{LeRoux2008SAPA09b,wisdom2018differentiable} is implicitly enforced when using the STFT.%

\subsection{Results}

Results on the universal data are shown in Figure \ref{fig:sisnr_vs_stftwin} and Table \ref{tab:divsource}, and audio demos may be found online \cite{supplemental_webpage}.
Figure \ref{fig:sisnr_vs_stftwin} shows results for different window sizes, where for each size, the hop is half the window size.  For comparison, speech/non-speech separation performance on data described in \cite{wisdom2018differentiable} is shown\footnote{Note that we consider here the more general ``speech/non-speech separation'' task, in contrast to the ``speech enhancement'' task, which typically refers to separating only the speech signal.} alongside results for two-source and three-source universal sound separation. We also tried training CLDNN networks with learned bases, but these networks failed to converge and are not shown. For all tasks, we show the performance of an oracle binary mask using an STFT for varying window sizes. These oracle scores provide a theoretical upper bound on the possible performance of our methods.

The differences between tasks in terms of basis type are striking. Notice that for speech/non-speech separation, longer STFT windows are preferred for all masking networks, while shorter windows are best when using a learnable basis. For universal sound separation, the optimal window sizes are shorter in general compared to speech/non-speech separation, regardless of the basis.

Window size is an important variable since it controls the frame rate and temporal resolution of the network, as well as the basis size in the case of STFT analysis and synthesis transforms.  The frame rate also determines the temporal context seen by the network.  On the speech/non-speech separation task, for all masking networks, 25-50 ms is the best window size.  Speech may work better with such relatively long windows for a variety of reasons: speech is largely voiced and has sustained harmonic tones, with both the pitch and vocal tract parameters varying relatively slowly. Thus, speech is well described by sparse patterns in an STFT with longer windows as preferred by the models, and may thus be easier to separate in this domain.  Speech is also highly structured and may carry more predictable longer-term contextual information than arbitrary sounds; with longer windows, the LSTM in a CLDNN has to remember information across fewer frames for a given temporal context.   

For universal sound separation, the TDCNs prefer short (2.5 ms or 5 ms) frames, and the optimal window size for the CLDNN is 5 ms or less, which in both cases is much shorter than the optimal window size for speech/non-speech separation.  This holds both with learned bases and with the STFT basis.   Surprisingly the STFT outperforms learned bases for sound separation overall, whereas the opposite is true for speech/non-speech separation. In contrast to speech/non-speech separation, where a learned basis can exploit the structure of speech signals, it is perhaps more difficult to learn general-purpose basis functions for the wide variety of acoustic patterns present in arbitrary sounds.
In contrast to speech, arbitrary sounds may contain more percussive components, and hence be better represented using an STFT with finer time resolution.
To fairly compare different models, we report results using the optimal window size for each architecture, determined via cross-validation.  

Table \ref{tab:divsource} shows summary comparisons using the best window size for each masking network and basis. The optimal performance for speech/non-speech separation is achieved by models using learnable bases, while for universal sound separation, STFTs provide a better representation. 
For both two-source and three-source separation, the iTDCN++ with 2.5 ms STFT basis provides the best average SI-SDR improvement of 
9.8 dB and
8.7 dB, respectively, on the test set, whereas the 2xTDCNN++, is not competitive on the universal separation task.
For speech/non-speech separation, the iTDCN++ with a 2.5 ms learnable basis achieves the best performance of 
13.4 dB SI-SDRi.
The iterative networks were trained with the loss function applied to the output of each iteration.
These results point to iterative separation as a promising direction for future exploration.

\begin{figure}[th]
    \centering
    \includegraphics[width=0.49\columnwidth]{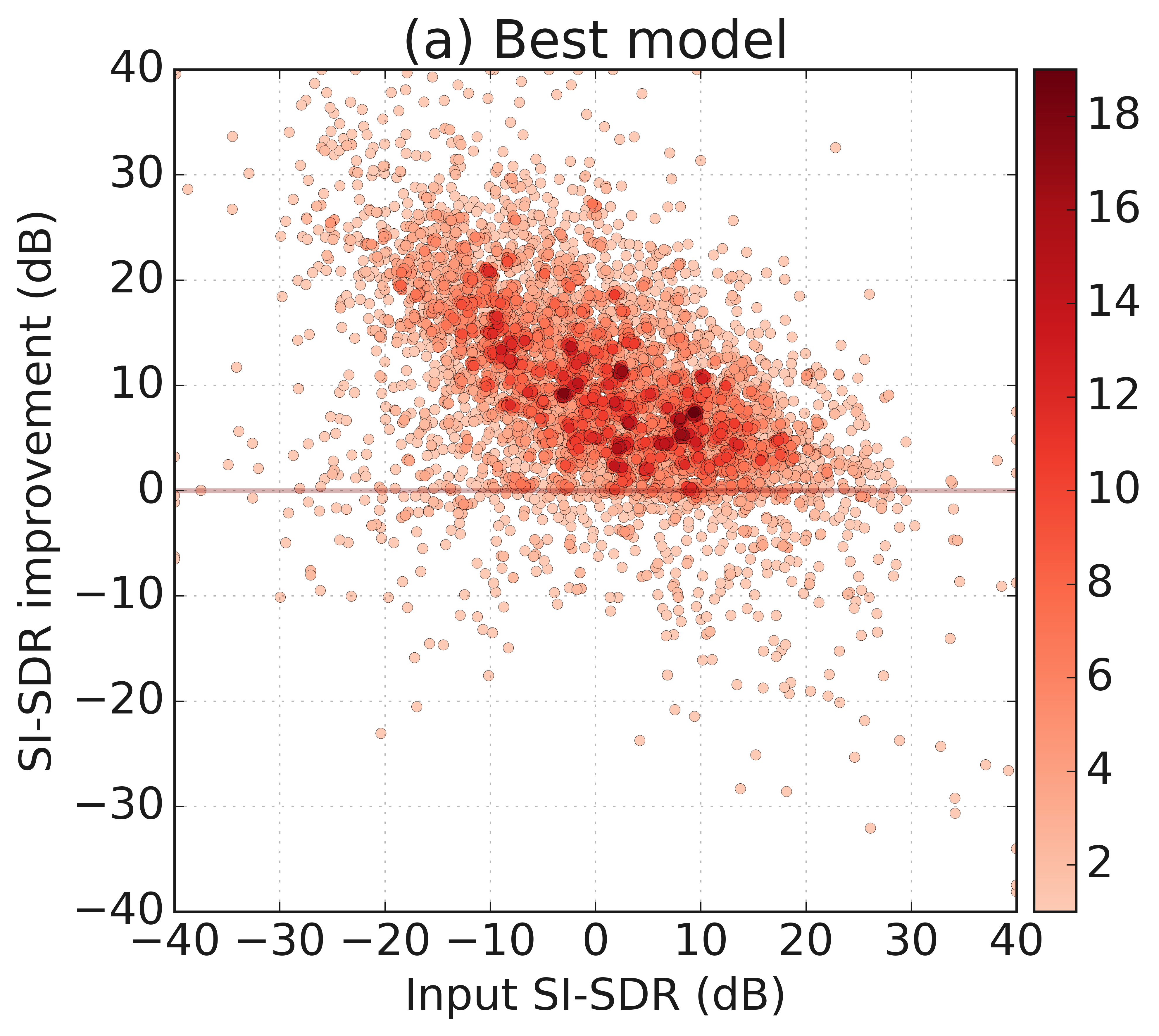}
    \includegraphics[width=0.49\columnwidth]{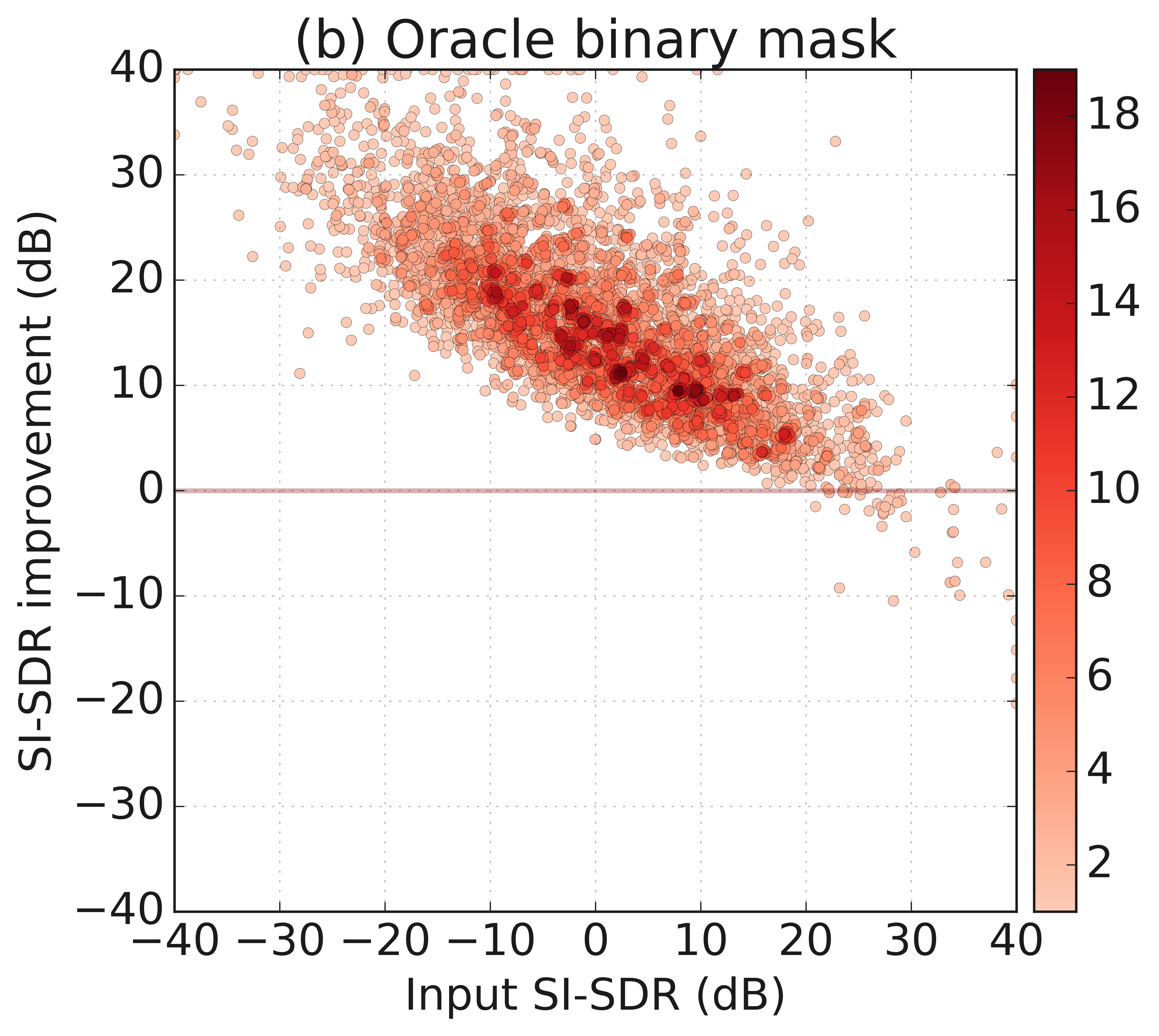}
    \vspace{-5pt}
    \caption{Scatter plots of input SI-SDR versus SI-SDR improvement on two-source universal sound mixture test set for a) our best model (iTDCN++, STFT) and b) oracle binary masking using an STFT with 10 ms window and 5 ms hop. The darkness of points is proportional to the number of overlapping points.}
    \label{fig:scatters}
    \vspace{-15pt}
\end{figure}

Figure \ref{fig:scatters} shows scatter plots of input SI-SDR versus improvement in SI-SDR  for each example in the test set. Panel a) displays results for the best model from Table \ref{tab:divsource}, and panel b) displays results for oracle binary masking computed using an STFT with 10 ms windows and 5 ms hop. Oracle binary masking achieves 
16.3 dB mean SI-SDRi, and indicates the potential separation that can be achieved on this dataset.

\section{Conclusion}
\label{sec:conc}

We introduced the universal sound separation problem and constructed a dataset of mixtures containing a wide variety of different sounds.  Our experiments compared different combinations of network architectures and analysis-synthesis transforms, optimizing each over the effect of window size. We also proposed novel variations in architecture, including longer-range skip-residual connections and iterative processing, that improve separation performance on all tasks.  Interestingly, the optimal basis and window size are different when separating speech versus separating arbitrary sounds, with learned bases working better for speech/non-speech separation, and STFTs working better for sound separation.  The best models, using iterative TDCN++, produce an average SI-SDR improvement of almost 10 dB on sound separation, and over 13 dB on speech/non-speech separation. Overall, these are extremely promising results which show that perhaps the holy grail of universal sound separation may soon be within reach. %

\vfill
\clearpage
\balance
\bibliographystyle{IEEEtran_nourl}
\bibliography{refs}

\begin{thebibliography}{10}
\def\url#1{}
\csname url@samestyle\endcsname
\providecommand{\newblock}{\relax}
\providecommand{\bibinfo}[2]{#2}
\providecommand{\BIBentrySTDinterwordspacing}{\spaceskip=0pt\relax}
\providecommand{\BIBentryALTinterwordstretchfactor}{4}
\providecommand{\BIBentryALTinterwordspacing}{\spaceskip=\fontdimen2\font plus
\BIBentryALTinterwordstretchfactor\fontdimen3\font minus
  \fontdimen4\font\relax}
\providecommand{\BIBforeignlanguage}[2]{{%
\expandafter\ifx\csname l@#1\endcsname\relax
\typeout{** WARNING: IEEEtran.bst: No hyphenation pattern has been}%
\typeout{** loaded for the language `#1'. Using the pattern for}%
\typeout{** the default language instead.}%
\else
\language=\csname l@#1\endcsname
\fi
#2}}
\providecommand{\BIBdecl}{\relax}
\BIBdecl

\bibitem{lu2013speech}
X.~Lu, Y.~Tsao, S.~Matsuda, and C.~Hori, ``Speech enhancement based on deep
  denoising autoencoder,'' in \emph{Proc. Interspeech}, Mar. 2013.

\bibitem{Weninger2014RNN}
F.~J. Weninger, J.~R. Hershey, J.~{Le Roux}, and B.~Schuller,
  ``Discriminatively trained recurrent neural networks for single-channel
  speech separation,'' in \emph{Proc. GlobalSIP}, Dec. 2014.

\bibitem{xu2014experimental}
Y.~Xu, J.~Du, L.-R. Dai, and C.-H. Lee, ``An experimental study on speech
  enhancement based on deep neural networks,'' \emph{IEEE Signal Processing
  Letters}, vol.~21, no.~1, 2014.

\bibitem{erdogan2015phase}
H.~Erdogan, J.~R. Hershey, S.~Watanabe, and J.~Le~Roux, ``Phase-sensitive and
  recognition-boosted speech separation using deep recurrent neural networks,''
  in \emph{Proc. ICASSP}, Apr. 2015.

\bibitem{weninger2015speech}
F.~Weninger, H.~Erdogan, S.~Watanabe, E.~Vincent, J.~Le~Roux, J.~R. Hershey,
  and B.~Schuller, ``Speech enhancement with {LSTM} recurrent neural networks
  and its application to noise-robust {ASR},'' in \emph{Proc. LVA/ICA}, Aug.
  2015.

\bibitem{Hershey2016ICASSP03}
J.~R. Hershey, Z.~Chen, J.~Le~Roux, and S.~Watanabe, ``Deep clustering:
  Discriminative embeddings for segmentation and separation,'' in \emph{Proc.
  ICASSP}, Mar. 2016.

\bibitem{Isik2016Interspeech09}
Y.~Isik, J.~Le~Roux, Z.~Chen, S.~Watanabe, and J.~R. Hershey, ``Single-channel
  multi-speaker separation using deep clustering,'' in \emph{Proc.
  Interspeech}, Sep. 2016.

\bibitem{Yu2017PIT}
D.~Yu, M.~Kolb{\ae}k, Z.-H. Tan, and J.~Jensen, ``Permutation invariant
  training of deep models for speaker-independent multi-talker speech
  separation,'' in \emph{Proc. ICASSP}, Mar. 2017.

\bibitem{kolbaek2017uPIT}
M.~Kolb{\ae}k, D.~Yu, Z.-H. Tan, J.~Jensen, M.~Kolbaek, D.~Yu, Z.-H. Tan, and
  J.~Jensen, ``Multitalker speech separation with utterance-level permutation
  invariant training of deep recurrent neural networks,'' \emph{IEEE/ACM
  Transactions on Audio, Speech and Language Processing}, vol.~25, no.~10,
  2017.

\bibitem{Wang2017Overview}
D.~Wang and J.~Chen, ``{Supervised Speech Separation Based on Deep Learning: An
  Overview},'' in \emph{arXiv preprint arXiv:1708.07524}, 2017.

\bibitem{Wang2018ICASSP04Alternative}
Z.-Q. Wang, J.~Le~Roux, and J.~R. Hershey, ``Alternative objective functions
  for deep clustering,'' in \emph{Proc. ICASSP}, Apr. 2018.

\bibitem{leroux2018sdr}
J.~Le~Roux, S.~Wisdom, H.~Erdogan, and J.~Hershey, ``{SDR} -- half-baked or
  well done?'' \emph{Proc. ICASSP}, May 2019.

\bibitem{huang2014deep}
P.-S. Huang, M.~Kim, M.~Hasegawa-Johnson, and P.~Smaragdis, ``Deep learning for
  monaural speech separation,'' in \emph{Proc. ICASSP}, May 2014.

\bibitem{luo2018tasnet}
Y.~Luo and N.~Mesgarani, ``Tasnet: Surpassing ideal time-frequency masking for
  speech separation,'' \emph{arXiv preprint arXiv:1809.07454}, 2018.

\bibitem{unet_spot}
A.~Jansson, E.~Humphrey, N.~Montecchio, R.~Bittner, A.~Kumar, and T.~Weyde,
  ``Singing voice separation with deep {U}-{N}et convolutional networks,''
  \emph{Proc. ISMIR}, Oct. 2017.

\bibitem{gan}
Y.~C. Subakan and P.~Smaragdis, ``Generative adversarial source separation,''
  in \emph{Proc. ICASSP}, Apr. 2018.

\bibitem{plumbley_latest}
E.~M. Grais and M.~D. Plumbley, ``Combining fully convolutional and recurrent
  neural networks for single channel audio source separation,'' in \emph{Proc.
  AES}, May 2018.

\bibitem{plumbley}
Q.~Kong, Y.~Xu, W.~Wang, and M.~D. Plumbley, ``A joint
  separation-classification model for sound event detection of weakly labelled
  data,'' in \emph{Proc. ICASSP}, Apr. 2018.

\bibitem{mmdenselstm}
N.~Takahashi, N.~Goswami, and Y.~Mitsufuji, ``{MMDenseLSTM}: An efficient
  combination of convolutional and recurrent neural networks for audio source
  separation,'' \emph{arXiv preprint arXiv:1805.02410}, 2018.

\bibitem{cdae}
E.~M. Grais and M.~D. Plumbley, ``Single channel audio source separation using
  convolutional denoising autoencoders,'' in \emph{Proc. GlobalSIP}, Nov. 2017.

\bibitem{onetomany}
P.~Chandna, M.~Miron, J.~Janer, and E.~G{\'o}mez, ``Monoaural audio source
  separation using deep convolutional neural networks,'' in \emph{Proc.
  LVA/ICA}, Feb. 2017.

\bibitem{chime2}
E.~Vincent, J.~Barker, S.~Watanabe, J.~{Le Roux}, F.~Nesta, and M.~Matassoni,
  ``The second `{CHiME}' speech separation and recognition challenge: Datasets,
  tasks and baselines,'' in \emph{Proc. ICASSP}, May 2013.

\bibitem{attractor}
Y.~Luo, Z.~Chen, and N.~Mesgarani, ``Speaker-independent speech separation with
  deep attractor network,'' \emph{IEEE/ACM Transactions on Audio, Speech, and
  Language Processing}, vol.~26, no.~4, 2018.

\bibitem{dpcl_music}
Y.~Luo, Z.~Chen, J.~R. Hershey, J.~Le~Roux, and N.~Mesgarani, ``Deep clustering
  and conventional networks for music separation: Stronger together,'' in
  \emph{Proc. ICASSP}, Mar. 2017.

\bibitem{wilson2018exploring}
K.~Wilson, M.~Chinen, J.~Thorpe, B.~Patton, J.~Hershey, A.~R. Saurous,
  J.~Skoglund, and F.~R. Lyon, ``Exploring tradeoffs in models for low-latency
  speech enhancement,'' in \emph{Proc. IWAENC}, Sep. 2018.

\bibitem{wisdom2018differentiable}
S.~Wisdom, J.~R. Hershey, K.~Wilson, J.~Thorpe, M.~Chinen, B.~Patton, and R.~A.
  Saurous, ``Differentiable consistency constraints for improved deep speech
  enhancement,'' \emph{Proc. ICASSP}, May 2019.

\bibitem{zhang2019fixup}
H.~Zhang, Y.~N. Dauphin, and T.~Ma, ``Fixup initialization: Residual learning
  without normalization,'' \emph{arXiv preprint arXiv:1901.09321}, 2019.

\bibitem{prosound}
\BIBentryALTinterwordspacing
Pro {S}ound {E}ffects {L}ibrary. Available from
  {http://www.prosoundeffects.com}, accessed: 2018-06-01.
  \url{http://www.prosoundeffects.com}
\BIBentrySTDinterwordspacing

\bibitem{supplemental_webpage}
\BIBentryALTinterwordspacing
Universal {S}ound {S}eparation project webpage. Available from
  {https://universal-sound-separation.github.io}.
  \url{https://universal-sound-separation.github.io}
\BIBentrySTDinterwordspacing

\bibitem{tf_short}
\BIBentryALTinterwordspacing
M.~A. et~al., ``{TensorFlow}: Large-scale machine learning on heterogeneous
  systems,'' 2015, software available from tensorflow.org.
  \url{http://tensorflow.org/}
\BIBentrySTDinterwordspacing

\bibitem{adam}
D.~P. Kingma and J.~Ba, ``Adam: A method for stochastic optimization,''
  \emph{arXiv preprint arXiv:1412.6980}, 2014.

\bibitem{LeRoux2008SAPA09b}
J.~{{Le} Roux}, N.~Ono, and S.~Sagayama, ``Explicit consistency constraints for
  {STFT} spectrograms and their application to phase reconstruction,'' in
  \emph{Proc. SAPA}, Sep. 2008.

\end{thebibliography}

\end{document}